\documentclass[twocolumn,twoside,slac_two]{revtex4}
\usepackage{graphicx}
\usepackage{fancyhdr}
\begin{document}

\title{New Heavy Quark Baryons}

%

\author{M. Kreps on behalf of the {\sl BABAR},
Belle and CDF Collaborations}
\affiliation{Institut f\"ur Experimentelle Kernphysik,
University of Karlsruhe, Germany}

\begin{abstract}
During the past year many interesting results were published
in heavy quark baryon spectroscopy. In addition to several
refined measurements, new states were
directly observed both in the charm and the bottom sector. In this
paper we review recent results on heavy quark baryons from 
B-factories and Tevatron experiments.
\end{abstract}

\maketitle

\thispagestyle{fancy}


\section{Introduction}

Heavy quark baryons provide, in the same way as heavy quark
mesons,
an interesting laboratory for studying and testing Quantum
Chromodynamics (QCD), the theory of strong interactions. The heavy
quark mesons provide the closest analogy to the hydrogen atom,
which provided important tests of Quantum Electrodynamics.
In this analogy we can consider the heavy quark meson as the
"hydrogen atom" of QCD. The heavy quark baryons are the next step,
where we have a state with one heavy quark and two light
quarks, which are often treated together as diquark and so
effectively providing the same laboratory as heavy quark mesons. 
The heavy quark states test regions of the QCD, where 
perturbation calculations cannot be
used and many different approaches to solve the theory were
developed. Just a few examples of them are Heavy quark
effective theory, non-relativistic and relativistic
potential models or lattice QCD.

\begin{figure}[bt]
\begin{center}
 \includegraphics[width=7.0cm]{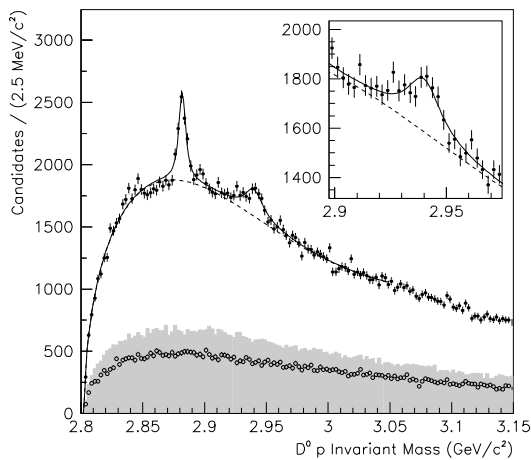}
 \caption{Invariant mass distribution of $pD^0$ from
the {\sl BABAR} experiment \cite{Aubert:2006sp}. The full points show
the experimental data and the open points represent wrong sign
($p\overline{D}^0$) combinations. The shaded area shows data
from $D^0$ sidebands. The solid curve is 
the result of the fit, with the dashed line showing the background 
part of the fit. 
\label{fig1}}
\end{center}
\end{figure}
From the experimental point of view, we approached a point, where
several experiments collected large data samples to
study the heavy quark baryons in detail, using fully
reconstructed decays. The recent results come from both
the asymmetric $e^+e^-$ B-factories with the {\sl BABAR} and
Belle experiments and the $p\overline{p}$ collisions at Tevatron
with CDF and D\O{}. The $e^+e^-$ B-factories profit from
huge datasets collected in a very clean environment, but their
deficit is that the energy is not high enough to study
b-quark baryons and are therefore limited to the charm sector. At
the Tevatron on the other hand all b-quark hadrons are
produced, but the price which has to be payed is a more
complicated environment with huge background coming from
the beam fragmentation.

In the past year many new states were observed. In addition several
measurements of the properties of already known states were
performed \cite{Abe:2006xr, Aubert:2007bt, Aubert:2006qw,
Aubert:2006rv}. 
 While all of the results are certainly
interesting, for space reasons we will concentrate here
only on the newly observed states, which are
$\Lambda_c(2940)$, $\Xi_c(2980)$ and $\Xi_c(3077)$, $\Omega_c^*$,
$\Sigma_b$ and $\Sigma_b^*$.

\section{$\Lambda_c$ states}

The $\Lambda_c$ is lowest lying baryon state in the charm sector.
Together with the ground state, four other states are listed
by the Particle Data Group \cite{Yao:2006px}. From these states,
the one with highest mass is $\Lambda_c(2880)$ seen by the CLEO experiment
\cite{Artuso:2000xy}. As this is the first $\Lambda_c$ state
above the $pD^0$ threshold, the {\sl BABAR}
Collaboration  performed a search for the
$\Lambda_c(2880)$ in the $p\overline{D}^0$ final state. The
resulting invariant mass distribution is shown in Fig.
\ref{fig1} \cite{Aubert:2006sp}. Together with the 
clear signal of $\Lambda_c(2880)$
state another resonant structure is seen at a mass
of $2.94\,\mathrm{GeV}/c^2$.
The parameters of the resonances are extracted using an unbinned maximum 
likelihood fit, where each of the two signals is
described by a relativistic Breit-Wigner function convoluted
with a Gaussian resolution function. The product of a fourth-order
polynomial and two-body phase space is used for the background
description. The
obtained values are listed in Table \ref{tab1}. The
significance of the newly observed $\Lambda_c(2940)^+$ is
$7.5$ standard deviations.
\begin{table*}[th]
\begin{center}
\tabcolsep=4mm
\caption{Measured masses and widths of the new $\Lambda_c$
and $\Xi_c$ states.}
\begin{tabular}{lcccc} \hline
  & \multicolumn{2}{c}{{\sl BABAR}} & \multicolumn{2}{c}{Belle} \\ 
State & Mass $[\mathrm{MeV}/c^2]$ & Width $[\mathrm{MeV}/c^2]$ & Mass $[\mathrm{MeV}/c^2]$ & Width $[\mathrm{MeV}/c^2]$ \\ \hline
$\Lambda_c(2880)^+$ & $2881.9\pm 0.1\pm 0.5$ & $5.8\pm 1.5\pm 1.1$ & $2881.2\pm 0.2\pm 0.4$ & $5.5\pm0.7\pm1.1$ \\
$\Lambda_c(2940)^+$ & $2939.8\pm  1.3\pm  1.0$ & $17.5\pm  5.2\pm  5.9$ & $2938.0\pm 1.3^{+2.0}_{-4.0}$ & $13\,^{+8}_{-5}\, ^{+27}_{-7}$ \\ \hline
$\Xi_c(2980)^+$ & $2967.1\pm1.9\pm1.0$ & $23.6\pm2.8\pm1.3$ &  $2978.5\pm2.1\pm2.0$ & $43.5\pm7.5\pm7.0$ \\
$\Xi_c(2980)^0$ & --- & --- & $2977.1\pm8.8\pm3.5$ & $43.5$ (fixed) \\
$\Xi_c(3077)^+$ & $3076.4\pm0.7\pm0.3$ & $6.2\pm1.6\pm0.5$ & $3076.7\pm0.9\pm0.5$ & $6.2\pm1.2\pm0.8$ \\ 
$\Xi_c(3077)^0$ & --- & --- & $3082.8\pm1.8\pm1.5$ & $5.2\pm3.1\pm1.8$ \\ \hline
\end{tabular}
\label{tab1}
\end{center}
\end{table*}

\begin{figure}[tb]
\begin{center}
 \begin{picture}(550,170)
\put(20,80){\rotatebox{90}{${\rm Events}\;/\; 2.5\,\mathrm{MeV}/c^2$}}
\put(80,0){$M(\Lambda_c\pi^+\pi^-),\,\mathrm{GeV}/c^2$}
\put(10,-20){\includegraphics[width=8.0cm]{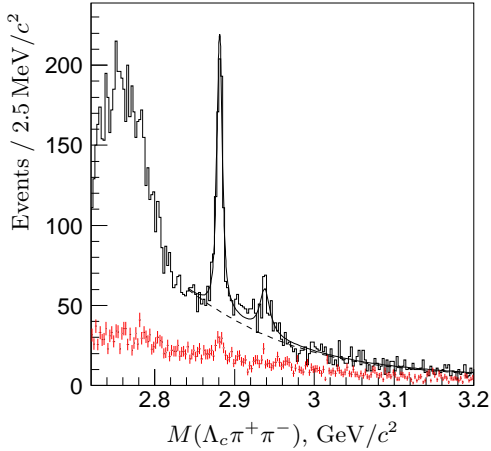}}
\end{picture}
 \caption{$\Lambda_c^+\pi^+\pi^-$ invariant mass
distribution from the Belle experiment \cite{Abe:2006rz}.
The histogram represents data with $\Lambda_c^+\pi^\pm$ being
consistent with $\Sigma_c(2455)$, red dots with error bars
show the scaled $\Sigma_c(2455)$ sideband distribution. The full and dashed
lines represent the result of the fit and the background part of the fit.
\label{fig2}}
\end{center}
\end{figure}

A confirmation of the newly observed state is reported by
the Belle
experiment in the $\Lambda_c^+\pi^+\pi^-$ final state. The
invariant mass distribution with requiring the 
$\Lambda_c^+\pi^\pm$ invariant mass to be consistent with
the $\Sigma_c(2455)$ is shown in Fig. \ref{fig2}
\cite{Abe:2006rz}. Three signals corresponding to
$\Lambda_c(2765)^+$, $\Lambda_c(2880)^+$ and
$\Lambda_c(2940)^+$ are clearly visible in
the distribution. Using a binned maximum likelihood fit with
two signals and a third-order polynomial background the
masses and widths of the $\Lambda_c(2880)^+$ and
$\Lambda_c(2940)^+$ states are extracted. The resulting parameters are listed
in Table \ref{tab1} and are consistent with those measured
by the {\sl BABAR} experiment. The significance of the
$\Lambda_c(2940)^+$ state is found to be $6.2$ standard
deviations.

\begin{figure}[tb]
\begin{center}
 \includegraphics[width=7.0cm]{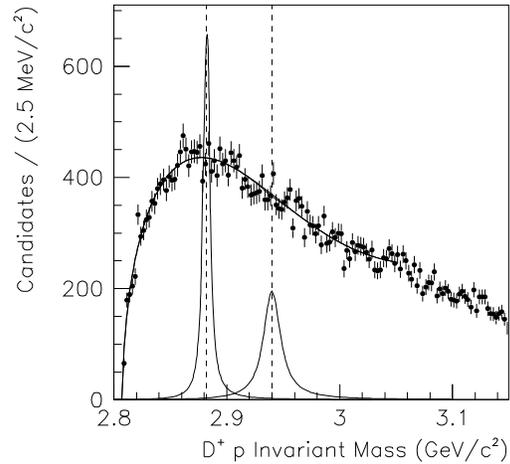}
 \caption{$pD^+$ invariant mass distribution from
the {\sl BABAR} experiment \cite{Aubert:2006sp}. The points with error bars
show data, the curve through the points is the result of the fit. The two
peaks show the expected signal if the two states at $2880$ and
$2940$ $\mathrm{MeV}/c^2$ would be produced with the same rate
as in the $pD^0$ final state. 
\label{fig3}}
\end{center}
\end{figure}

To gain more information on the two $\Lambda_c$ states, both
experiments perform additional studies. A first question to
which the experiments try to find an answer is whether the $\Lambda_c(2940)^+$
and $\Lambda_c(2880)^+$ are really $\Lambda_c$ states and
not $\Sigma_c$ states. In the case when the observed states are $\Sigma_c$ states,
one also expects signals in the $pD^+$ invariant mass
distribution. {\sl BABAR} performs this study, the result
of which is shown in Fig. \ref{fig3}. None of the two
signals are visible from which one concludes, that the
observed states are $\Lambda_c$ states. For illustration,
the two peaks in Fig. \ref{fig3} show expected signals if the states would
be $\Sigma_c$ states produced with the same rate as in the $pD^0$
channel. 

\begin{figure}[tb]
\centering
\begin{picture}(550,320)
\put(18,260){\rotatebox{90}{${\rm Events}\;/\; 0.2$}} 
\put(18,85){\rotatebox{90}{${\rm Events}\;/\; (\pi/10)$}} 
\put(120,163){$\cos\theta$} 
\put(120,2){$\phi$, rad} 
\put(10,-20){\includegraphics[width=8cm]{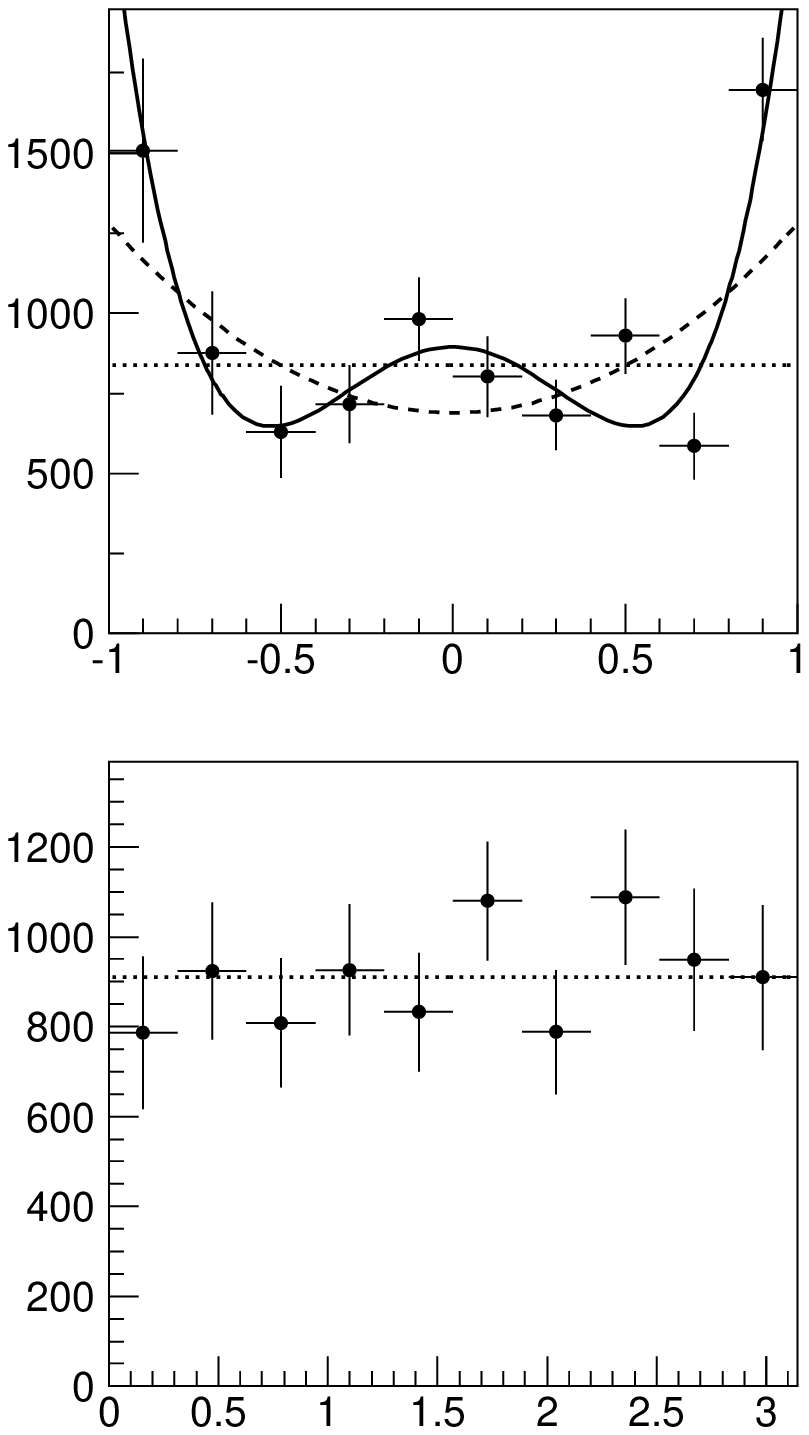}}
\end{picture}
\caption{The yield of $\Lambda_c(2880)\rightarrow\Sigma_c\pi$ decays
as a function of $\cos\theta$ and $\phi$. The lines shows
the fit for $J=1/2$ (dotted), $J=3/2$ (dashed) and $J=5/2$
(full) hypothesis.}
\label{fig4a}
\end{figure}

\begin{figure}[tb]
\begin{center}
 \begin{picture}(550,170)
 \put(20,87){\rotatebox{90}{${\rm Events}\;/\; 2\,\mathrm{MeV}/c^2$}}
\put(90,0){$M(\Lambda_c^+\pi^{\pm}),\,\mathrm{GeV}/c^2$}
\put(10,-20){ \includegraphics[width=8.0cm]{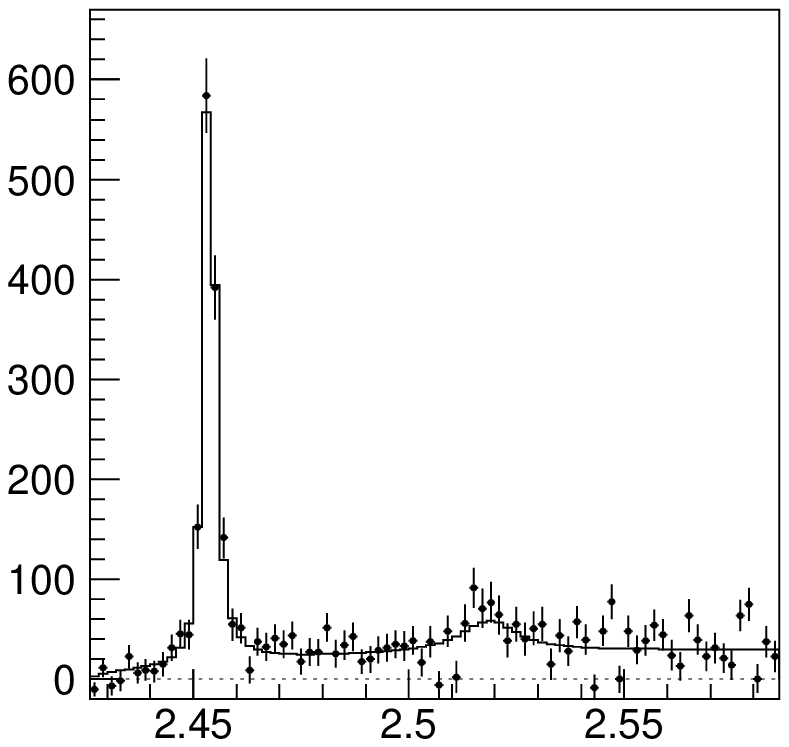}}
\end{picture}
 \caption{The $\Lambda_c(2880)^+$ yield as a function of
the invariant mass of $\Lambda_c^+\pi^\pm$. The histogram shows
the result of the fit.
\label{fig4}}
\end{center}
\end{figure}

The Belle experiment uses a different approach and uses a high
statistics signal for the $\Lambda_c(2880)^+$ state to study
the resonant
substructure of its decay to $\Lambda_c^+\pi^+\pi^-$. Also
an analysis of the angular distributions is done to constrain
the $\Lambda_c(2880)^+$ quantum
numbers. To gain more statistics, these studies are done 
with looser cuts compared to the ones used for 
the invariant mass distribution of Fig. \ref{fig2}. 
To obtain the sub-resonant structure of the
decay, fits of $\Lambda_c^+\pi^+\pi^-$ invariant mass
distributions in slices of the invariant mass
$M(\Lambda_c^+\pi^\pm)$ are done. The obtained distributions of
$\Lambda_c(2880)^+$ events is
shown in Fig. \ref{fig4}. A clear signal for
the $\Sigma_c(2455)$ is visible together with a structure at
the mass
of the $\Sigma_c(2520)$ state. The significance of the
$\Sigma_c(2520)$ signal is estimated to be $3.7$ standard
deviations. The resulting ratios of partial widths are
$\frac{\Gamma(\Sigma_c(2455)\pi^\pm)}{\Gamma(\Lambda_c^+\pi^+\pi^-)}=0.404\pm0.021\pm0.014$,
$\frac{\Gamma(\Sigma_c(2520)\pi^\pm)}{\Gamma(\Lambda_c^+\pi^+\pi^-)}=0.091\pm0.025\pm0.010$
and
$\frac{\Gamma(\Sigma_c(2520)\pi^\pm)}{\Gamma(\Sigma_c(2455)\pi^\pm)}=0.225\pm0.062\pm0.025$.
The angular analysis uses the helicity angle $\theta$ and
the angle $\phi$ between the
$e^+e^-\rightarrow \Lambda_c(2880)^+ X$ reaction plane and
a plane defined by the pion momentum and the $\Lambda_c(2880)^+$ boost
direction in the $\Lambda_c(2880)^+$ rest frame. The
measured angular distributions are shown in Fig.
\ref{fig4a}.%
The distribution of $\phi$ is found to be uniform, which
is consistent with a $J=1/2$ hypothesis. The $J=3/2$ and
$J=5/2$ hypotheses, which are considered as well, can have a uniform
$\phi$ distribution but doesn't have to. In the description
of the $\cos\theta$
distribution for all three hypotheses only terms consistent
with a flat $\phi$ distributions are used. By a $\chi^2$
comparison it is concluded, that the most probable assignment is
$J=5/2$ and that the two lower spins can be excluded by $5.5$ or $4.8$
standard deviations. Having determined the spin of the state, we can return
to the measurement of
$R=\frac{\Gamma(\Sigma_c(2520)\pi^\pm)}{\Gamma(\Sigma_c(2455)\pi^\pm)}$.
 Heavy quark symmetry predicts
$R=1.4$ for a $5/2^-$ state and $R=0.23$ --- $0.36$ for a $5/2^+$
state. Comparing the prediction to the measured value, Belle
concludes that the $\Lambda_c(2880)^+$ state has
quantum numbers $J^P=5/2^+$.

Here we would like to point out, that a state at $2765$
$\mathrm{MeV}/c^2$ is listed by the Particle Data Group
\cite{Yao:2006px}, but
its nature is unknown. It would certainly be important to
perform more studies of this state to find out, whether it
is a $\Lambda_c$ or $\Sigma_c$ excitation and to perform an angular
analysis to constrain the quantum numbers of this state.
Certainly both experiments at the B-factories should have enough
data to perform such an analysis.

\section{$\Xi_c$ states}

\begin{figure}[tb]
\begin{center}
 \includegraphics[width=8.0cm]{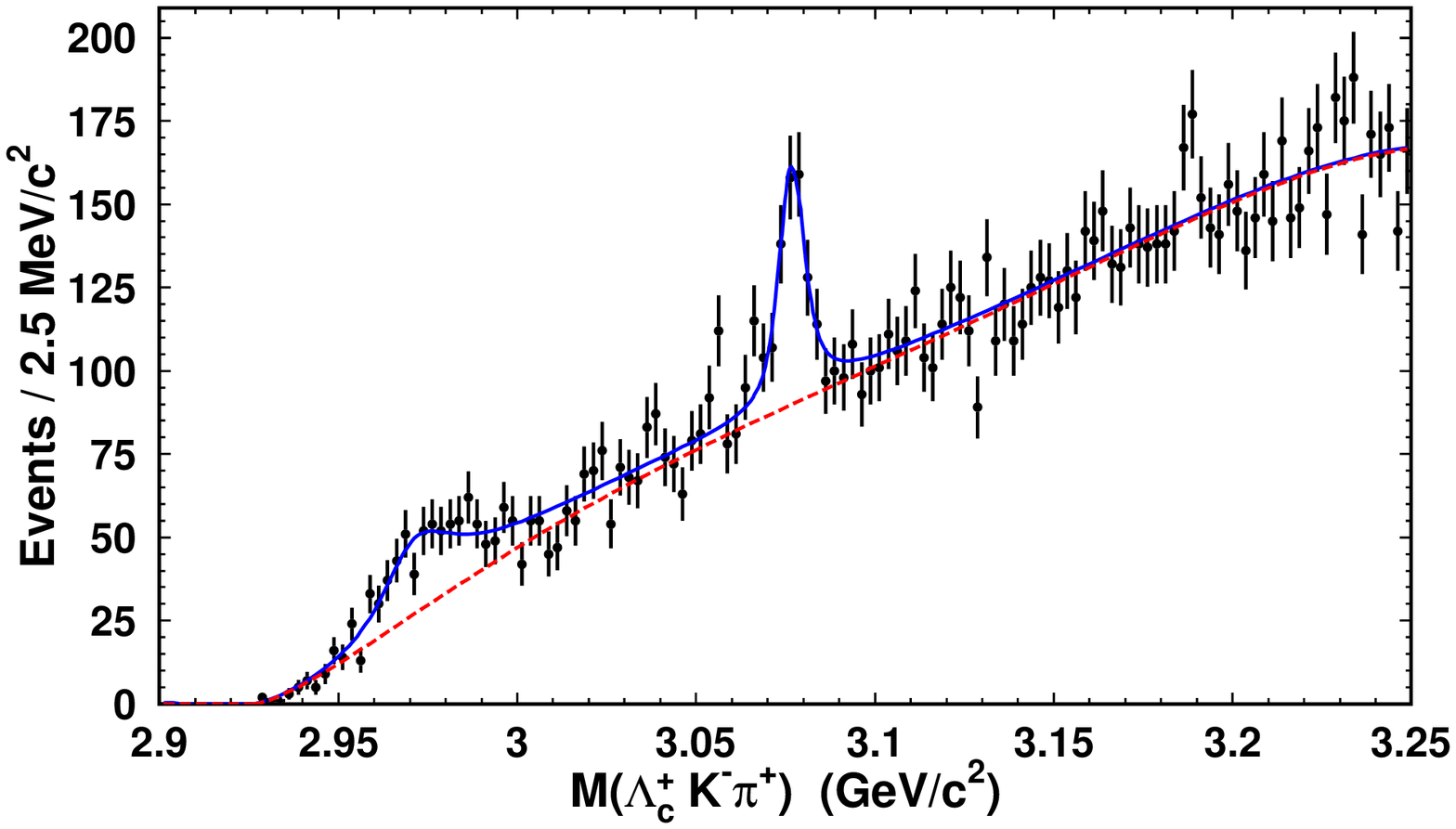}
 \includegraphics[width=8.0cm]{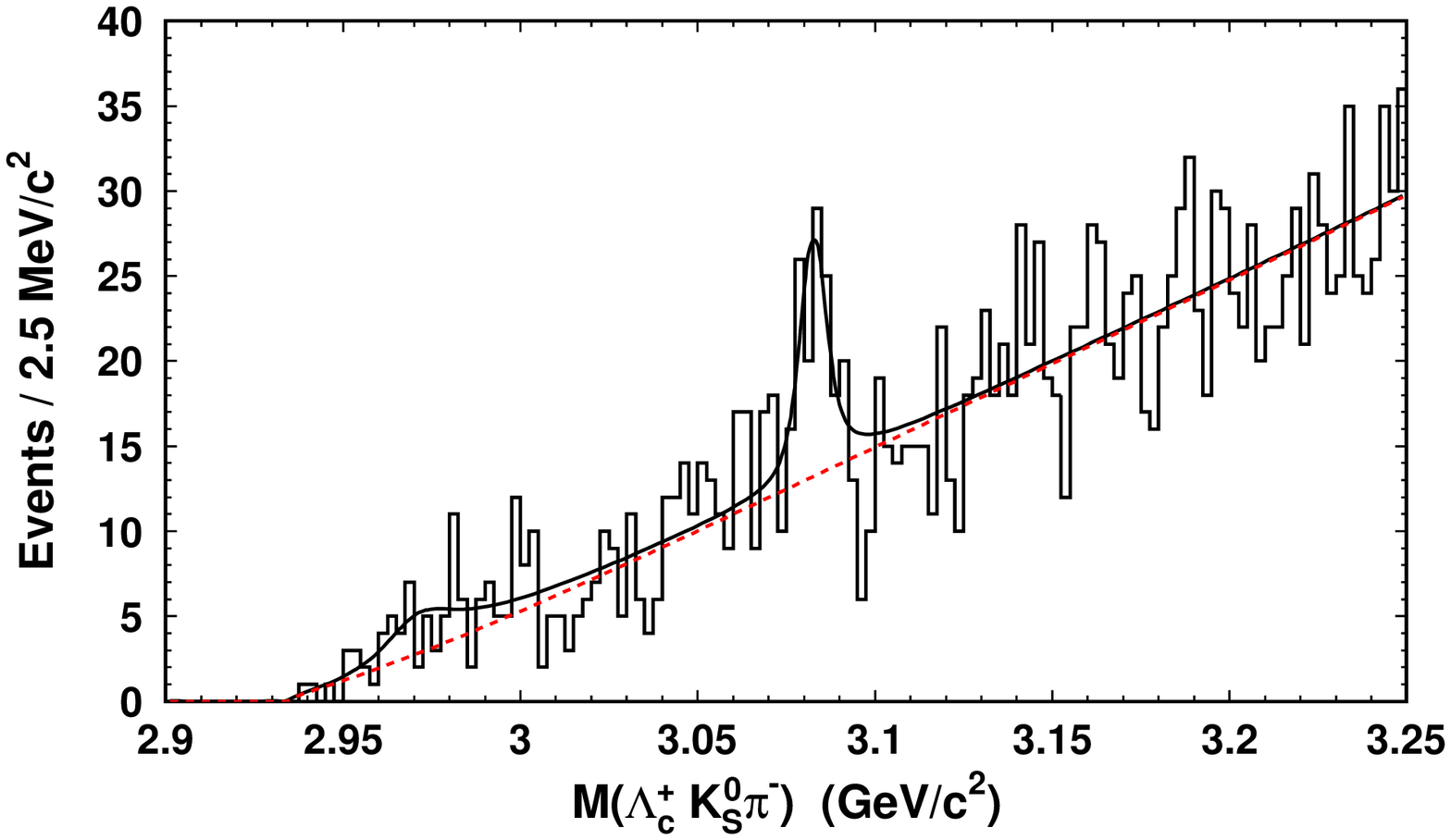}
 \caption{Distribution of the invariant mass distribution of
$\Lambda_c^+K^-\pi^+$ combinations (top) and
$\Lambda_c^+K_s^0\pi^+$ combinations (bottom)
\cite{Chistov:2006zj}. The points
represent data, the full (blue) line corresponds to the fit result
and the dashed (red) line to the background part of the fit.
\label{fig5}}
\end{center}
\end{figure}

Another particle for which new states are observed during
the last
year is the $\Xi_c$. One motivation is the observation
of the double charmed baryon $\Xi_{cc}$ in the $\Lambda_c^+K^-\pi^+$
final state reported by the SELEX collaboration
\cite{Mattson:2002vu}. The other
is that charmed strange baryons should decay to this final
state if they are above threshold, but none of the states
seen up to now have high enough mass.
In this situation the Belle collaboration performs a search for
the new baryons in the $\Lambda_c^+K^-\pi^+$ and
$\Lambda_c^+K_s^0\pi^+$ final states. The invariant mass
distributions in both decay modes are shown in Fig.
\ref{fig5} \cite{Chistov:2006zj}.
Two clear peaks are visible in the $\Lambda_c^+K^-\pi^+$ final
state and one in the $\Lambda_c^+K_s^0\pi^+$ final state. From
an maximum likelihood fit the masses, widths, numbers of events
and significances are extracted. The fit returns
$405.3\,\pm\,50.7$ signal events for the $\Xi_c(2980)^+$ state
and $326.0\,\pm\,39.6$ signal events for the $\Xi_c(3077)^+$.
The significances of the two states are $5$ and $9$ standard
deviations. In the case of the neutral combination the fit results in
$67.1\,\pm\,19.9$ events for the $\Xi_c(3077)^0$ and
$42.3\,\pm\,23.8$ events for the $\Xi_c(2980)^0$. Due to the low
statistics the width of 
the $\Xi_c(2980)^0$ state was fixed to be the same as for the charged state.
The significance of the $\Xi_c(3077)^0$ is $4$ standard
deviations, while the statistical significance of
the $\Xi_c(2980)^0$ state is $1.5$ standard deviations. The
obtained masses and widths are listed in Table \ref{tab1}. 
The interpretation of the newly observed states as $\Xi_c$'s
can unambiguously be derived
from the quark content of the particles in the final state,
which gives the quark content of a $\Xi_c$ baryon.

\begin{figure}[tb]
\begin{center}
 \includegraphics[width=8.0cm]{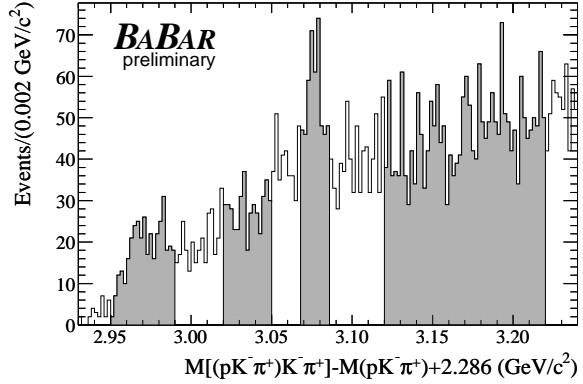}
 \caption{Distribution of the invariant mass difference of
$\Lambda_c^+K^-\pi^+$ from the {\sl BABAR} experiment
\cite{Aubert:2006uw}. The shaded regions were used for dalitz
plots, which can be found in the original work
\cite{Aubert:2006uw}.
\label{fig6}}
\end{center}
\end{figure}
The observation of the $\Xi_c(2980)^+$ and $\Xi_c(3077)^+$
states is confirmed by the {\sl BABAR} experiment using
the $\Lambda_c^+K^-\pi^+$ decay mode \cite{Aubert:2006uw}. The invariant mass
difference distribution is shown in Fig. \ref{fig6}.
Structures at the masses of the two new $\Xi_c$ states seen
by Belle are also visible here. The {\sl BABAR}
collaboration doesn't exploit neutral charge combinations,
but rather goes further and studies the resonance substructure
of the decays of the $\Xi_c(2980)^+$ and $\Xi_c(3077)^+$ states.
\begin{figure}[tb]
\begin{center}
 \includegraphics[width=8.0cm]{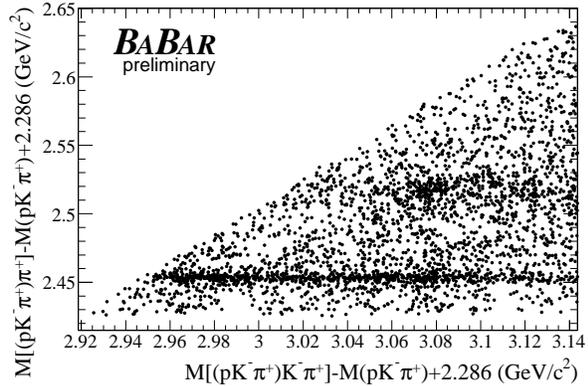}
 \caption{A two-dimensional distribution of $M(\Sigma_c)$
versus $M(\Xi_c)$ used in the fit to extract $\Xi_c$ states
parameters. The two horizontal bands correspond to
the $\Sigma_c(2455)^{++}$ and $\Sigma_c(2520)^{++}$ states.
\label{fig7}}
\end{center}
\end{figure}
\begin{figure}[tb]
\begin{center}
 \includegraphics[width=8.0cm]{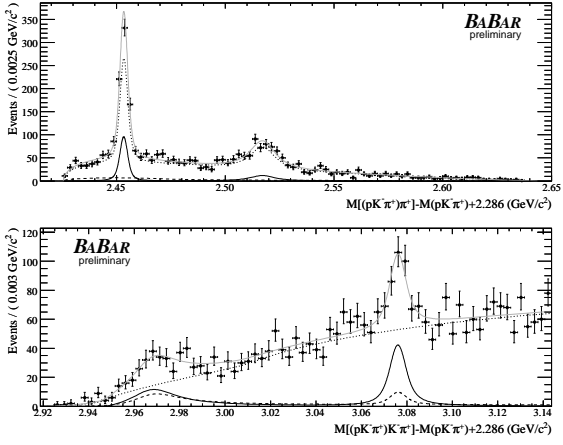}
 \caption{Projections of the fit onto the mass variables
$M(\Lambda_c^+\pi^+)$
(top) and $M(\Lambda_c^+K^-\pi^+)$ (bottom). The points represent
data and the solid grey curves the full fit. The signal component with
the $\Sigma_c^{++}$ intermediate states are shown by the solid
dark curve and the non-resonant part of the decay by the dashed
curve. The fitted background is shown by the dotted curve.
\label{fig8}}
\end{center}
\end{figure}
In the first step, the dalitz plots of $M(\pi^+K^-)$ versus
$M(\Lambda_c^+\pi^+)$ in the four shaded regions
in Fig. \ref{fig6} are examined. In the dalitz distributions for
which we kindly refer reader to work \cite{Aubert:2006uw}, a clear
structure corresponding to the $\Sigma_c(2455)^{++}$ and
the $\Sigma_c(2520)^{++}$ are visible in both signal and
sideband regions. In the final extended maximum likelihood
fit of $\Lambda_c^+K^-\pi^+$ and $\Lambda_c^+\pi^+$ masses
the amount of decays through $\Sigma_c(2455)^{++}$,
$\Sigma_c(2520)^{++}$ and non-resonant $\Lambda_c^+K^-\pi^+$
is extracted. The two dimensional distribution of
$M(\Lambda_c^+\pi^+)$ versus $M(\Lambda_c^+K^-\pi^+)$ used in the fit is shown in
Fig. \ref{fig7}.  Figure \ref{fig8} shows fit projections
onto $M(\Lambda_c^+\pi^+)$ and $M(\Lambda_c^+K^-\pi^+)$.
The fit returns $284\,\pm\,45\,\pm\,46$ signal events for
the $\Xi_c(2980)^+$ with a significance of $7$ standard deviations
and $204\,\pm\,35\,\pm\,12$ signal events with a
significance of $8.6$
standard deviations for the $\Xi_c(3077)^+$. The extracted masses and
widths are listed in Table \ref{tab1}. Results for the
$\Xi_c(3077)^+$ agree well between Belle and {\sl
BABAR}, while for the $\Xi_c(2980)^+$ there is a slight
discrepancy between the two experiments.
\begin{table}
\caption{\label{tab2} Yields and significances for the
separate resonant and non-resonant decays of the $\Xi_c(2980)^+$
and  $\Xi_c(3077)^+$ states \cite{Aubert:2006uw}.}
\begin{center}
\begin{tabular}{lcc} \hline
 & Events     & Significance     \\ \hline
$\Xi_c(2980)^+\rightarrow\Sigma_c(2455)^{++}K^-$     &
$132\pm31\pm5$     & $4.9\,\sigma$    \\
$\Xi_c(2980)^+\rightarrow\Lambda_c^+K^-\pi^+$        &
$152\pm37\pm45$    & $4.1\,\sigma$    \\ \hline
$\Xi_c(3077)^+\rightarrow\Sigma_c(2455)^{++}K^-$     &
$87\pm20\pm4$      & $5.8\,\sigma$    \\
$\Xi_c(3077)^+\rightarrow\Sigma_c(2520)^{++}K^-$     &
$82\pm23\pm6$      & $4.6\,\sigma$    \\
$\Xi_c(3077)^+\rightarrow\Lambda_c^+K^-\pi^+$        &
$35\pm24\pm16$     & $1.4\,\sigma$    \\ \hline
\end{tabular}
\end{center}
\end{table}
The yields and significances of different resonant and
non-resonant decays are listed in Table \ref{tab2}. Except
of the non-resonant component of $\Xi_c(3077)^+$ decay, all
others are in the range from $4$ to $6$ standard deviations.
For both new $\Xi_c$ states the decays through
$\Sigma_c(2455)$ and $\Sigma_c(2520)$ resonances forms 
a large part of the observed decays.

For the future, both states can benefit from further studies. On
the one hand, a confirmation on $5$ standard deviations
level for the 
neutral states should be done. The other direction
clearly is to establish spin and parity of the new states.

\section{Observation of the $\Omega_c^*$ state}

\begin{figure}[htb]
\begin{center}
 \includegraphics[width=8.0cm]{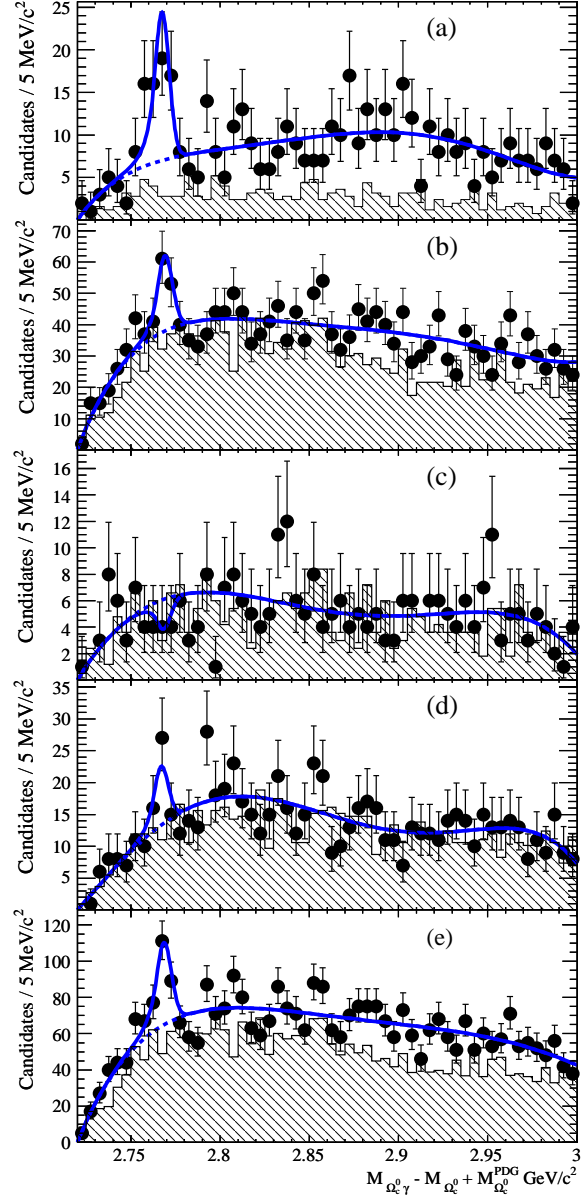}
 \caption{The invariant mass difference distributions of
$\Omega_c^*\,\rightarrow\,\Omega_c\gamma$ candidates, with
the $\Omega_c$ reconstructed in (a) $\Omega^-\pi+$, (b)
$\Omega^-\pi^+ \pi^0$, (c) $\Omega^-\pi^+\pi^-\pi^+$, (d) 
$\Xi^-K^-\pi^+\pi^+$ decay mode and (e) combining all decay
modes together.
The points with error bars represent data and the dashed
line distribution from the $\Omega_c$ sidebands. The full line
represents the result of the fit and the dashed line the combinatorial
background.
\label{fig9}}
\end{center}
\end{figure}
\begin{table*}[tbh]
\caption{The mass difference
$\Delta{M}=M(\Omega_c^*)-M(\Omega_c)$, the fitted signal yield
$Y$ (events) and the $\Omega_c^*$ signal significance using
different $\Omega_c$ decay modes.  }
\begin{center}
\tabcolsep=4mm
\renewcommand{\arraystretch}{1.4}
\begin{tabular}{ l c c c c } \hline
Decay mode& $\Delta M$ ($\mathrm{MeV}/c^2$) & $Y$ (Events) & $S$ ($\sigma$)  \\ \hline
 $\Omega_c^0 \rightarrow \Omega^-\pi^+$  & $69.9 \pm 1.4 \pm 1.0$ & $39 ^{+10}_{-9} \pm 6$& 4.2  \\
 $\Omega_c^0 \rightarrow \Omega^-\pi^+ \pi^0$  & $71.8 \pm 1.3 \pm 1.1$ & $55^{+16}_{-15}\pm 6$& 3.4\\
 $\Omega_c^0 \rightarrow \Omega^-\pi^+\pi^-\pi^+$  &  69.9 (fixed) & $-5\pm 5 \pm 1$ & -\\
 $\Omega_c^0 \rightarrow \Xi^-K^-\pi^+\pi^+$  & $69.4^{+1.9}_{-2.0} \pm 1.0$ & $20 \pm 9 \pm 3$ & 2.0 \\
Combined  & $70.8 \pm 1.0 \pm1.1$ & $105 \pm 21 \pm6$ & 5.2 \\ \hline
\end{tabular}
\end{center}
\label{tab3}
\end{table*}
\renewcommand{\arraystretch}{1.0}
In the charm sector, all singly charmed states with zero
orbital momentum have been discovered \cite{Yao:2006px} except
the $\Omega_c^*$. The theoretical expectations for the mass
difference between the $\Omega_c^*$ and the $\Omega_c$ are in the
range from $50$ to $104$ $\mathrm{MeV}/c^2$
\cite{Mathur:2002ce,Woloshyn:2000qa, Burakovsky:1997vm,
Savage:1995dw, Rosner:1995yu, Roncaglia:1995az,
Lichtenberg:1995kg, Zalewska:1996az, Glozman:1995xy, Jenkins:1996de}. 
In the work of the {\sl BABAR}
experiment \cite{Aubert:2006je} a search for the  $\Omega_c^*$
through its radiative decay is performed. The $\Omega_c$ is
reconstructed in the following decay modes
\begin{eqnarray}
  \Omega_c^0 \rightarrow \Omega^-\pi^+ &\quad&
  \Omega_c^0 \rightarrow \Omega^-\pi^+ \pi^0 \nonumber \\
  \Omega_c^0 \rightarrow \Omega^-\pi^+\pi^-\pi^+ &\quad&
  \Omega_c^0 \rightarrow \Xi^-K^-\pi^+\pi^+ \nonumber 
\end{eqnarray}
with $\Omega^-\,\rightarrow\,\Lambda K^-$ and
$\Xi^-\,\rightarrow\,\Lambda \pi^-$. In total around $300$ $\Omega_c$
signal events are reconstructed in all four decay modes.
The $\Omega_c$ candidates
are then combined with a photon to form $\Omega_c^*$
candidates. The invariant mass difference distributions in
the four different $\Omega_c$ decay modes are shown in Fig.
\ref{fig9}.
In the two decay modes with the highest $\Omega_c$ signal, a clear
peak in the $\Omega_c\gamma$ invariant mass difference
distribution is visible, see Fig. \ref{fig9}(a) and
\ref{fig9}(b). In the other two modes one mode has
an indication for a signal, while the other one has not. The
maximum likelihood fit in each $\Omega_c$ decay mode yields
consistent mass differences
$\Delta{M}=M(\Omega_c^*)-M(\Omega_c)$ and widths across the
$\Omega_c$ decay modes (see Table
\ref{tab3}). As the separate decay modes are consistent, we can
combine them together and perform a single maximum
likelihood fit. The result of the fit is shown in Fig. \ref{fig9}(e)
and yields $\Delta
M\,=\,70.9\,\pm\,1.0\,\pm\,1.1\,\mathrm{MeV}/c^2$ with
$105\,\pm\,21\,\pm\,6$ signal events. The significance of the
signal including systematic uncertainty is $5.2$ standard
deviations.

\section{Observation of charged $\Sigma_b$ and $\Sigma_b^*$}

Up to recently, the $\Lambda_b$ was the only directly
observed $b$-baryon. With increasing data samples collected at the Tevatron
accelerator, the searches for other b-baryons starts to be
feasible. The first of such searches was performed by the CDF experiment,
which searched for the $\Sigma_b$ baryon and its spin
excited partner $\Sigma_b^*$ \cite{Sigma_b}. A general theoretical
expectations \cite{Mathur:2002ce, Jenkins:1996de,
Stanley:1980fe, Stanley:1980zm, Izatt:1981pt,
Lichtenberg:1989an, Martin:1986qt, Richard:1983tc,
Basdevant:1985ux, Bowler:1996ws, Jenkins:1996rr,
Albertus:2003sx, Ebert:2005xj, Hwang:1986ee,
Capstick:1987cw, Kwong:1987mj, Karliner:2003sy} 
are the mass difference
$M(\Sigma_b)-M(\Lambda_b)-M(\pi)\,=\,40\,$--$\,70\,\mathrm{MeV}/c^2$
with
$M(\Sigma_b^*)-M(\Sigma_b)\,=\,10\,$--$\,40\,\mathrm{MeV}/c^2$.
A small difference on the level of $5$ $\mathrm{MeV}/c^2$ is
expected between the masses of $\Sigma_b^+$ and $\Sigma_b^-$.
Both the $\Sigma_b$ and the $\Sigma_b^*$ are expected to be narrow
with a natural width of around $8$ and $15$ $\mathrm{MeV}/c^2$
with $\Lambda_b \pi$ being the dominant decay mode.

The CDF search is based on $1$ $fb^{-1}$ of data using
fully reconstructed $\Lambda_b$ baryons. $The \Lambda_b$ is reconstructed in
the $\Lambda_c\pi$ decay mode with
$\Lambda_c\,\rightarrow\,pK^-\pi^+$. In total around $3000$
$\Lambda_b$ signal events are reconstructed. In the sample
used for the $\Sigma_b$ search $86$ \% of events are $\Lambda_b$
baryons. The search is performed for the charged
$\Sigma_b$'s
only, as the neutral one decays by emission of $\pi^0$,
which is extremely difficult to detect at the CDF
experiment.

\begin{figure}[tb]
\begin{center}
 \includegraphics[width=8.0cm]{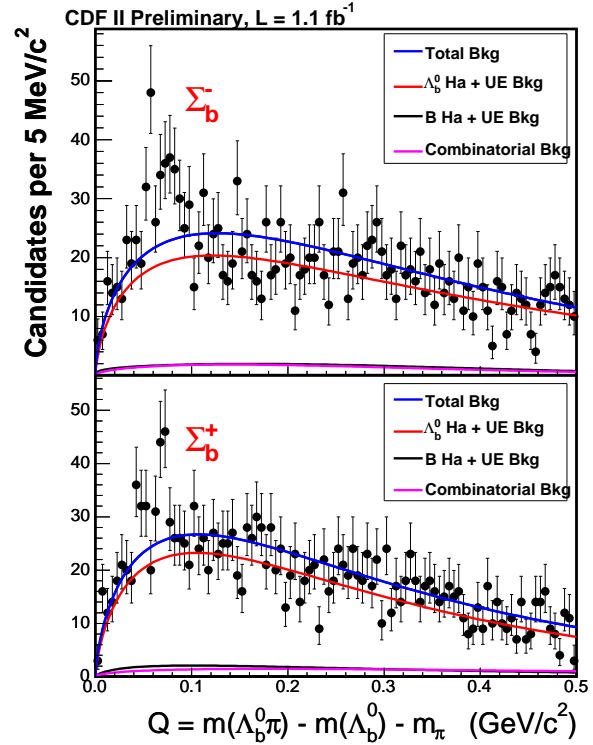}
 \caption{The invariant mass difference distribution of
the $\Sigma_b^\pm$ candidates. The points with error bars
represent data, the blue line represents the predicted
background, while the other three lines show three separate background
contributions.
\label{fig11}}
\end{center}
\end{figure}
The selected $\Lambda_b$ candidates are then combined with
charged pions to form $\Sigma_b$ candidates. After fixing
the selection of candidates, the background is estimated while
keeping the signal region blinded. The background consists of
three basic components, which are combinatorial background,
$\Lambda_b$ hadronization and hadronization of
mis-reconstructed $B$ mesons. Relative fractions of these
components are taken from the fit of the $\Lambda_b$ invariant mass
distribution. The shape of the combinatorial background is
determined using the upper sideband of the $\Lambda_b$
invariant mass
distribution. For the hadronization of mis-reconstructed $B$
mesons the fully reconstructed
$B^0\,\rightarrow\,D^-\pi^+$ in the data are used. The shape of the
largest component, $\Lambda_b$ hadronization, is determined
using a \textsc{pythia} Monte Carlo sample. The unblinded mass
difference distributions
with the predicted background are shown in Fig. \ref{fig11}.
For the both charges a clear excess above the predicted background is
visible in the signal region
($Q\,\in\,[30,100]\,\mathrm{MeV}/c^2$).
The number of background events predicted in the signal region
is $268$ for $\Sigma_b^-$ and $298$ for $\Sigma_b^+$. In the data we observe
$416$ events for negatively charged candidates and $406$ for
positively charged ones.  

\begin{figure}[tb]
\begin{center}
 \includegraphics[width=8.0cm]{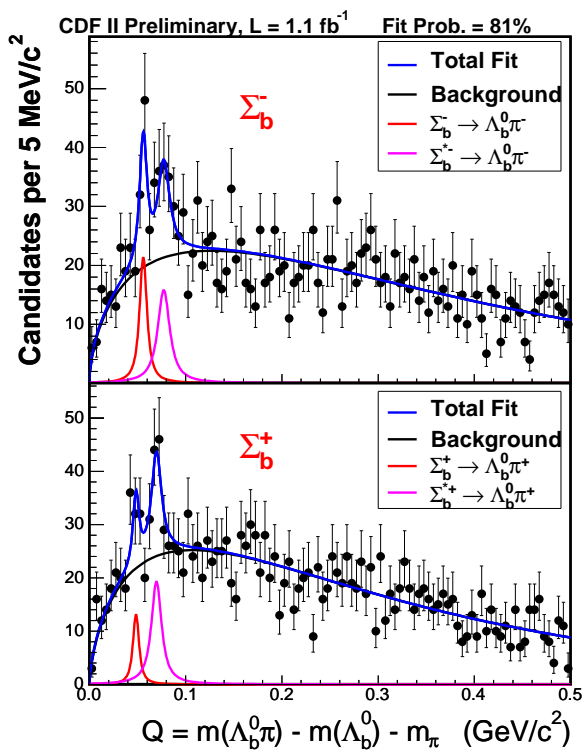}
 \caption{Projection of the fit result of the $\Sigma_b$
invariant mass difference distribution. The points with
error bars represent data. The blue line corresponds to the result of
the fit, the background is shown by the black line while the
signals are represented by the
red and magenta curves.
\label{fig12}}
\end{center}
\end{figure}
\begin{table}[tb]
\caption{\label{tab4}Result of the fit to the $\Sigma_b$
invariant mass difference distribution. } 
\tabcolsep=4mm
\begin{center}
\renewcommand{\arraystretch}{1.4}
\begin{tabular}{lc} \hline
 Parameter & Value \\ \hline
$Q(\Sigma_b^+)$ ($\mathrm{MeV}/c^2$) & $48.4^{+2.0}_{-2.3}\pm 0.1$ \\
$Q(\Sigma_b^-)$ ($\mathrm{MeV}/c^2$) & $55.9^{+1.0}_{-1.0}\pm 0.1$ \\
$M(\Sigma_b^{*})-M(\Sigma_b)$ ($\mathrm{MeV}/c^2$) & $21.3^{+2.0}_{-1.9}{} ^{+0.4}_{-0.2}$ \\ \hline
$\Sigma_b^+$ events & $29^{+12.4}_{-11.6}{} ^{+5.0}_{-3.4}$ \\
$\Sigma_b^-$ events & $60^{+14.8}_{-13.8}{} ^{+8.5}_{-4.0} $ \\
$\Sigma_b^{*+}$ events & $74^{+17.2}_{-16.3}{} ^{+10.3}_{-5.7}$ \\
$\Sigma_b^{*-}$ events & $74^{+18.2}_{-17.4}{} ^{+15.6}_{-5.0}$\\ \hline
\end{tabular}
\renewcommand{\arraystretch}{1.0}
\end{center}
\end{table}
To extract the signal yields and positions of the peaks,
an unbinned maximum likelihood fit is performed. The data are
described by a previously determined background shape
together with Breit-Wigner functions
convoluted with a resolution function for each peak. Due to
the low statistics, $M(\Sigma_b^{*+})\,-\,M(\Sigma_b^{+})$
is constrained to be the same as
$M(\Sigma_b^{*-})\,-\,M(\Sigma_b^{-})$.
The values obtained in the fit are summarized in Table
\ref{tab4} and the fit projection is shown in Fig. \ref{fig12}.

To estimate the significance of the observed signal, the fit is
repeated with alternative hypothesis and difference in the
likelihoods is used. Three different alternative hypotheses were
examined, namely the null hypothesis, using only two peaks instead of
four and leaving each single peak separately out of the fit. As a
result we conclude, that the null hypothesis can be excluded by
more than $5$ standard deviations. The fit also clearly
prefers four peaks against two and except of the $\Sigma_b^+$
peak, each peak has a significance above three standard
deviations.

\section{Conclusions}

The heavy quark baryons provide an interesting laboratory for
testing various approaches to the non-perturbative regime of
Quantum Chromodynamics. In the past year several new states
were discovered in the charm sector by the Belle and {\sl
BABAR} experiments and the charged $\Sigma_b$ and
$\Sigma_b^*$ baryons
in the bottom sector by the CDF experiment. The important point is that
experimentalists don't stop at the observation and the mass and
width measurements of the new states, but try to go beyond
to learn more. Certainly the angular analyzes to determine
the quantum numbers of the new states together with the detailed
studies of the production and decays is
important to learn more about heavy quark baryons in
general.

We conclude, that the last year was very productive in
experimental studies of the heavy quark baryons with several
important observations of new states. Probably the two most
important ones are the observation of the $\Omega_c^*$ and
the charged
$\Sigma_b^{(*)}$ baryons. With still increasing datasets and
current encouraging results we are convinced that more new
results can be expected in the upcoming year.

\begin{acknowledgments}
The author would like to thank all his colleagues from
the {\sl BABAR}, Belle and CDF experiments, who
contributed to the preparation of this talk and proceedings by
checking the material and giving useful comments.
\end{acknowledgments}

\bigskip 

\end{document}